\documentclass{iau} 
\usepackage{graphicx}

\title[The AstroCamp Project] 
{The AstroCamp Project}
\author[C. J. A. P. Martins]   
{C. J. A. P. Martins}
\affiliation{Centro de Astrof\'{\i}sica, Universidade do Porto, Rua das Estrelas, 4150-762 Porto, Portugal\\ email: {\tt Carlos.Martins@astro.up.pt}}

\pubyear{2021}
\volume{367}  
\jname{Education and Heritage in the era of Big Data in Astronomy}
\editors{R.M. Ros, B. Garcia, S. Gullberg, J. Moldon \& P. Rojo, eds.}
\begin{document}
\maketitle

\begin{abstract}
This contribution describes the concept, main structure and goals, and some highlighted outcomes, of the AstroCamp---an international academic excellence program in the field of astronomy and physics created in 2012 and organized by Centro de Astrof\'{\i}sica da Universidade do Porto (CAUP) together with the Paredes de Coura municipality and several national and international partners.
\keywords{Pre-university education, International programs, Academic excellence programs}
\end{abstract}

\firstsection 
\section{Introduction}

The AstroCamp is an academic excellence program in astronomy and physics created by the author in 2012 and organized by CAUP with the Paredes de Coura municipality and other partners---including, since 2017, the European Southern Observatory (ESO). It is intended for students in the last 3 years of pre-university education. Initially it was restricted to students living and studying in Portugal, but it now accepts applications from (in the 2020 and 2021 editions) 42 eligible countries, in Europe and the Americas.

The three main Astrocamp goals are to promote scientific knowledge, with high-quality training in a secluded and tranquil setting, to stimulate student curiosity and skills of critical thinking, team work and group responsibility, and to stimulate interactions between students with different backgrounds and life experiences but common interests.

AstroCampers are selected according to their motivation, academic merit and potential. As a point of principle the camp is free for students living in Portugal, and for foreign students the costs (if they exist) are a maximum of 400 Euro. In recent years about half of the applicants have been foreign. In common with other academic excellence programs, AstroCamp typically attracts more applications from girls than from boys.

\section{Structure and logistics}

The core of the AstroCamp is a two-week residential camp, held in mid-August at the Centre for Environmental Education of the Corno de Bico protected landscape, in the Paredes de Coura municipality (in the north of Portugal). Due to COVID-19 the 2020 edition was a hybrid one, with students living in Portugal in residence as usual and students living in other countries joining through a set on online collaboration tools.

Participation by invitation, after an application (in April) and a selection phase including an interview in English (at the end of May). There are no quotas of any kind, so academic merit and potential are the only selection criteria. In the 9 editions from 2012 to 2020 a total of 118 students (67 girls and 51 boys, from 14 countries) have been selected to participate. Basic statistics of the camp participants are in Table \ref{tab1}.

\begin{table}
  \begin{center}
  \caption{Basic statistics on 2012-2020 AstroCamp participants.}
  \label{tab1}
  \begin{tabular}{|c|c|c|}
  \hline 
{\bf Accepted Students} & {\bf Years 2012-2015} & {\bf Years 2016-2020}\\
\hline
Portuguese & $100\%$ & $54\%$ \\
Foreign & $0\%$ & $46\%$ \\
\hline
$10^{th}$ Grade & $24\%$ & $34\%$ \\
$11^{th}$ Grade & $52\%$ & $45\%$ \\
$12^{th}$ Grade & $24\%$ & $21\%$ \\
\hline
Boys & $47\%$ & $41\%$ \\
Girls & $53\%$ & $59\%$ \\
\hline
  \end{tabular}
 \end{center}
\vspace{1mm}
 {\it Notes:} In the first four editions, only Portuguese students were eligible. The middle part uses the Portuguese names for the last three years of pre-university education.
\end{table}

The core AstroCamp scientific activity are two courses, each with 10x90min classes and a two-hour written exam. Course lecturers must have a PhD and be currently active in research. (A pre-approved list of courses is proposed to the students, and they can then choose the ones they prefer to take.) Other activities are observational and high-level scientific programming projects (in Python or Matlab), stargazing sessions, Zoom chats with foreign researchers and selected documentaries. There are also community service projects and public talks, hiking (including an overnight one) and other recreational activities. Finally there are post-camp research and mentorship projects.

An example of a community service activity is the Solar System Hiking Trail. This is one of only 16 (in Europe) scales model of the Solar System, that is accurate both in terms of sizes and distances of the objects, which was developed during the first four AstroCamps, and officially opened on 13 August 2016.

The camp is fully residential: students, teachers and camp monitors (former AstroCamp students now doing university degrees) work and live together for 14 days. By choice the students have no internet access (and only have their mobile phones for very limited periods) to maximize their interactions with their peers and to enable them to focus on learning at a much faster pace than in their normal school classes.

\section{Highlights of Outcomes}

The scientific excellence of the camp can be illustrated by the fact that work done in the camp computational project has been included in several peer-reviewed publications, three recent examples being \cite[Alves et al. (2017)]{Alves17}, \cite[Alves et al. (2018)]{Alves18} and \cite[Faria et al. (2019)]{Faria19}. 

Post-camp student debriefings and follow-up activities (including an annual alumni lunch and an alumni weekend during the camp) demonstrate its impact. As an example, at the time of writing, 5 of the 10 students of the 2012 editions are doing PhDs in science topics, in the universities of Aveiro, Cambridge (two of them), Edinburgh and Minho.

\section*{Acknowledgments}

This work was funded by FEDER-COMPETE2020, and Portuguese FCT funds, under project POCI-01-0145-FEDER-028987 and PTDC/FIS-AST/28987/2017. The AstroCamp is supported by Funda\c c\~ao Millenium bcp, Ci\^encia Viva, U.Porto and ESO.

\end{document}